\documentclass{article}

\usepackage{microtype}
\usepackage{graphicx}
\usepackage{subcaption}
\usepackage{booktabs} 

\usepackage{hyperref}
\usepackage{xcolor}
\usepackage{soul}

\usepackage[accepted]{icml2026}

\usepackage{amsmath}
\usepackage{amssymb}
\usepackage{mathtools}
\usepackage{amsthm}

\usepackage[capitalize,noabbrev]{cleveref}

\theoremstyle{plain}

\theoremstyle{definition}

\theoremstyle{remark}

\usepackage[textsize=tiny]{todonotes}

\icmltitlerunning{The Foreign Policy AI Evaluation Gap}

\begin{document}

\twocolumn[
  \icmltitle{The Foreign Policy AI Evaluation Gap}
    
  \icmlsetsymbol{equal}{*}

  \begin{icmlauthorlist}
    \icmlauthor{Charles Pozniak}{equal,yyy}
    \icmlauthor{Jeba Sania}{equal,yyy}

  \end{icmlauthorlist}

  \icmlaffiliation{yyy}{Belfer Center, Harvard Kennedy School, Cambridge, MA, USA}

  \icmlcorrespondingauthor{Charles Pozniak}{cpozniak@hks.harvard.edu}
  \icmlcorrespondingauthor {Jeba Sania} {jebasania@hks.harvard.edu}

\icmlkeywords{technical AI governance, evaluation, statecraft, foreign policy, diplomacy, AI safety}
  \vskip 0.3in
]



\printAffiliationsAndNotice{}  

\begin{abstract}
We argue that AI systems used in conducting foreign policy tasks –– broadly enacting 'statecraft' -- should be a priority test case for technical AI governance research. In enacting foreign policy, we refer to the formulation and implementation of external objectives by political actors. Statecraft is a high-consequence deployment domain, with  extreme downside risks and structural properties that standard evaluation practices handle poorly. These features include partial observability, unbounded action spaces, contested ground truth, and multidimensional objectives. This paper advocates for a literature-grounded research agenda. Our contribution is threefold: (i) a claim about the structural conditions of foreign policy that combine catastrophic tail risk with technical evaluation complexities, (ii) an \textsc{ecosystem} review that highlights the asymmetric focus on \textsc{Assessment} features over \textsc{Access, Verification, Security}, and \textsc{Operationalization}, and (iii) a demand-side evaluation framework that decomposes foreign-policy workflows into bounded, evaluable sub-tasks with human recombination. As AI systems are already being deployed in the conduct of war and peace, amid limited public evaluation infrastructure from the technical AI governance community, this agenda is an urgent priority.
\end{abstract}

\section{Introduction}

AI systems have already entered foreign-policy workflows before the technical AI governance community has built evaluation infrastructure for them. Many initial uses appear administrative, but LLM systems are increasingly being explored for analysis, planning, and decision-support functions in military, diplomatic, and geopolitical contexts \cite{horowitz_bending_2023, rivera_escalation_2024, lamparth_human_2024, ma2024chatgptdonttelldo}. Technical AI governance must catch up. 

The technical AI governance literature provides a strong starting vocabulary. The Reuel--Bucknall taxonomy organizes the field around six governance capacities---\textsc{Assessment, Access, Verification, Security, Operationalization}, and \textsc{Ecosystem Monitoring}---as well as targets focused on \textsc{Data, Compute, Models}, and \textsc{Deployment} \citep{reuel_open_2025, bucknall_structured_2023}. The taxonomy is intentionally domain-general. Our claim is that foreign policy exposes a domain-specific gap, with capacities most needed where ordinary evaluation conditions are weakest.

This paper argues that foreign policy is a domain blind spot in technical AI governance. By conducting foreign policy or statecraft we mean the purposive, institutionally mediated conduct through which a political actor seeks to shape its external environment.\footnote{The paradigmatic actor is the state, but the definition can also include international organizations, coalitions, and de facto authorities when they formulate external objectives and act toward other political communities.} This includes both the formation and implementation of external objectives: diplomatic positions, negotiations, sanctions, crisis response, and the use of force. This definition is narrower than International Relations as a whole, because it concerns the policies and decision processes of actors rather than the entire structure of the international system, but it is broader than diplomacy, featuring a broader range of instruments, such as economic and legal tools, alongside negotiation and representation \cite{allison_essence_1999, halperin_bureaucratic_1974,putnam_diplomacy_1988, schelling_strategy_1960, raiffa_negotiation_2002, lax_3-d_2006}. 

We use foreign policy and statecraft to identify a diverse deployment domain, rather than a monolithic capability to be tested. Whether a model is ``good'' at foreign policy is somewhat irrelevant, instead this speaks to real-world human--AI workflow integrations in high-stakes domains.

For the technical AI governance audience, this focus on foreign policy is important for both methodological and normative reasons. World politics violates the assumptions under which benchmark practice is easiest, while also being a domain of great immediate consequence to human life. 

This paper makes three contributions. First, it explains why foreign policy is a structurally difficult evaluation domain, organized around representation, inference, and specification. Second, it maps these difficulties to the six TAIG governance capacities, showing that the gap spans the capacities. Third, it reports a corpus-screening study and external scope check that quantify the scarcity and \textsc{Assessment}-heavy character of existing foreign-policy AI evaluation work, then proposes a demand-side framework for task-scoped evaluation under institutional and strategic constraints.

The systems most in need of disciplined evaluation are also the systems least likely to yield clean benchmarks, public datasets, stable labels, or repeatable tests.

\section{Foreign Policy as a Priority Deployment Domain}

Though many domains are ``high stakes,'' foreign policy combines high consequence with poor measurability, strategic opacity, and weak external auditability. These factors mean that deployed systems evade effective technical governance while failures remain highly consequential. There are many instruments of foreign policy that vary across key dimensions of evaluability. Despite these differences, their institutional commonalities persist: activities ranging from arms control to crisis diplomacy are conducted through state action under strategic interaction.  This shared deployment setting creates recurring evaluation problems, while the appropriate evaluation design must remain task- and subdomain-specific.

Foreign policy is a ``high-stakes'' domain in which failure carries consequences of a categorically different order than those in commercial or private-sector settings. In matters of economic policy, diplomacy, and military decision-making, errors can be catastrophic and, in some cases, irreversible. Poor trade policy can destabilize economies, misread signals can escalate conflicts, and daily decisions determine life and death. States cannot reasonably rely on beta-testing during crises or delicate negotiations, and the nature of these stakes means that governments are often structurally unwilling to tolerate the degree of risk that iterative improvement requires. Similarly, states are unwilling to admit error for the same reasons. This risk is compounded by two features particularly acute in enacting foreign policy: (i) the use of highly destructive resources and (ii) the nature of externalities to citizens domestic and foreign. 

States, with their monopolies on violence, frequently deploy the most powerful and destructive resources. This includes both coercive force and ballistic weaponry, but also instruments like sanctions and export controls that are designed to inflict economic pain. An AI system that mishandles the application of these instruments, even marginally, operates in a domain where the consequences of error scale with the power of the actor deploying it. In no other setting is action taken with such a possibility for widespread destruction, both through physical and emotional harm and by limitations on the freedoms and capacities for self-actualization of individuals. 

In foreign policy, the actions of states impact many individuals outside of the internal decision processes and the national domestic democratic process, if one exists. These externalities are distinct from most private-sector failures. The impacts of errors in these settings are like with monopolies: the affected citizens may be unable to exercise accountability, move to another polity, or take remedial actions. A state-led AI failure in diplomacy can rapidly impose costs on large numbers of individuals, as well as on the institutions and norms of the international system. Failed conflict mediation between two states, for instance, can trigger refugee crises, disrupt regional security, and impose economic and political costs on states far removed from the original dispute. Existing governance regimes are not equipped to fully internalize these externalities. Unlike private sector deployments, this creates risks that neither markets nor regulatory frameworks can fully manage.

\section{Structural Limits of Evaluation}

There are several structural features of world politics that mean AI systems used for foreign policy can resist governance mechanisms. These classic properties of world politics result in evaluative difficulties across three key domains: \textbf{representation}, \textbf{inference}, and \textbf{specification}. 

Foreign policy settings are difficult to \textbf{represent} mathematically. Unlike in the board game \emph{Diplomacy}, the state space is unbounded. Though rules exist, the very institutions that enforce them can be dissolved through negotiations. The action space is similarly large: the range of packages, moves, proposals, legal instruments, and tactics is enormous \cite{raiffa_negotiation_2002}. Unlike individuals or profit-seeking businesses that are broadly bound under domestic law, foreign policy faces a different realm of constraints. Reductions of this complexity to single-shot interactions or multiple-choice options can mislead when treated as proxies for holistic, end-to-end judgment in diplomacy. They can, however, still be useful when scoped to bounded subtasks.

Contested ground truth is a constitutive feature of world politics, and one that makes accurate \textbf{inference} difficult.\footnote{In this setting, inference refers to inferring a true state or belief from noisy information. This differs from the classic ``inference'' of the AI stack.} In foreign policy, preferences are partly private, beliefs are noisy, and relevant counterfactual scenarios are unobservable \cite{fearon_rationalist_1995, jervis_perception_1976}. Strategies operate through commitment mechanisms, strategic signaling, and manipulation \cite{schelling_strategy_1960}. Strategic misrepresentation---of both actions and preferences---means that actors are also actively trying to limit the very ground truth on which effective action (and evaluation) depend.

Finally, success \textbf{specification} is an immediate challenge in foreign policy with multiple actors, each with their own preferences. Foreign-policy actors rarely optimize a single scalar objective. They may pursue security, legitimacy, economic advantage, and domestic political survival at the same time. Diplomatic decisions are also constrained by domestic ratification and bureaucratic process rather than by a single principal with a fixed utility function \citep{putnam_diplomacy_1988,allison_essence_1999,halperin_bureaucratic_1974}. Defining success is therefore partly a political act. Good evaluations should reward systems that surface trade-offs and preserve uncertainty, rather than forcing a single technical answer or collapsing ambiguity.

\subsection{Implications for Technical AI Governance}
These structural conditions make success difficult to evaluate, but they also have downstream implications for the AI governance \textsc{ecosystem}. The incentives for public disclosure are weak on both sides. Diplomatic and security operations are sensitive and states hesitate to publicize the settings in which they deploy AI systems. Similarly, commercial model developers face commercial and contractual pressures not to disclose failure cases or deployment constraints in classified, adversarial settings. The result is a domain in which the most consequential deployments are the least visible.

This lack of transparency creates dangerous information asymmetries. Sensitive deployments are least likely to be disclosed. Governments can become reliant on AI systems developed by private companies outside of the public safety discourse. States have strategic reasons to conceal the tasks, data, models, and evaluation standards used in diplomatic, intelligence, and security workflows. Vendors also have commercial and contractual reasons to say little. This is the inverse of much technical AI governance research, where the test set and results are  publicly available. 

The reliance on a small number of components in the stack means that strong one-time results are insufficient. Models are updated, benchmarks become gamed, and test items can leak into training data \cite{kapoor_ai_2025, liang_benchmarking_2025}. This opens up a narrow surface area of manipulability, as counterparts can exploit vulnerabilities. Shared vendors and common model families can also create correlated failure risks across ministries and allies \cite{horowitz_artificial_2018, rivera_escalation_2024, jensen2025criticalforeignpolicydecisions}. Foreign policy, therefore, requires longitudinal evaluation and diligent ecosystem-level monitoring. 

These dynamics are further destabilized by competitive pressures between states \cite{horowitz_artificial_2018}. Individually rational decisions to develop and deploy AI tools for strategic advantage can accelerate race dynamics, undermining international stability, eroding trust, and increasing the risk of crisis. Technological developments will continue to reconfigure the practice of diplomacy \cite{horowitz_artificial_2018, scharre_army_2018}. 

These barriers contribute toward a significant governance gap. AI use in geopolitical analysis is already a high-consequence domain, yet the existing literature on AI governance lacks a shared, externally inspectable evaluation practice that is commensurate with that consequence. Closing this gap requires not only better benchmarks, but also the institutional will from states and model developers to prioritize new evaluation methodologies, with coordination mechanisms to ensure transparency. 

\section{A TAIG Capacity Gap}
Despite the high stakes of deploying AI systems in foreign policy contexts, existing governance mechanisms remain inadequate across multiple dimensions. Technical AI governance can be understood as encompassing six interrelated subareas (\textsc{Assessment}, \textsc{Access}, \textsc{Verification}, \textsc{Security}, \textsc{Operationalization}, and \textsc{Ecosystem Monitoring}) \cite{reuel_open_2025}. In the following section, we highlight each capacity in relation to statecraft.

\textsc{Assessment} refers to the evaluation of AI systems for technical properties, risks, and broader second-order implications for society at large \cite{reuel_open_2025}. Currently, there is a lack of rigorous approaches to benchmark construction that reflect the complexity of real diplomatic environments, as well as a lack of expert validation drawn from practitioners to validate model outputs. Beyond this, well-known pitfalls of evaluation methods, such as avoiding model contamination, assessing adversarial perturbations, and calibration, also apply \cite{eriksson2025trustaibenchmarksinterdisciplinary, hutchinson2022evaluationgapsmachinelearning,shevlane_model_2023}. Additionally, there is a lack of evaluations focused on rare but catastrophic failure modes. 

\textsc{Access} refers to the ability to interact with AI systems and obtain relevant data and information while avoiding unreasonable infringement of privacy \cite{reuel_open_2025}. Currently, there are inadequate external access mechanisms to support the training and evaluation of AI systems used for foreign policy, leaving governments with limited ability to scrutinize the systems they deploy \cite{reuel_open_2025,raji2022outsideroversightdesigningparty}. Addressing this gap will require investments in infrastructure on both sides: governments need to build internal capacity to meaningfully engage with these systems, and developers need to provide mechanisms for external access. In practice, this might include secure access regimes to protect sensitive information, redacted system cards that disclose relevant model properties without compromising national security, synthetic scenarios that allow stress testing without revealing classified data, and audit sandboxes \cite{brundage2020trustworthyaidevelopmentmechanisms}. Without such access, meaningful oversight of AI systems will be relegated to the realm of theory.  

\textsc{Verification} is the ability of developers or third parties to verify claims made about AI systems' development, behavior, capabilities, or safety \cite{reuel_open_2025}. Unlike many commercial applications where outcomes can be observed and measured after deployment, AI systems used in foreign policy operate within complex, interdependent systems where ex-ante outcome verification is often structurally impossible. The consequences of a diplomatic decision may unfold over months or years, involve unintended actors, and resist neat attribution to any single choice or intervention. This makes traditional verification approaches, which tend to assume observable, discrete markers, ill-suited to the domain. Rather than attempting to verify outcomes after the fact, governance frameworks should prioritize verification at the process level. This means examining the intermediate steps and decision logic of AI systems in foreign policy before and during deployment, and decomposing complex tasks into components where claims about system behavior can be meaningfully assessed.

\textsc{Security} refers to the protection of AI system components from unauthorized access, use, or tampering \cite{reuel_open_2025}. This area takes on heightened significance in high-stakes domains such as foreign policy. Threat models in diplomatic and foreign-policy settings can be vastly more consequential than in commercial deployments in terms of scale and intensity. The cost of a security failure, whether through unauthorized access to model internals, manipulation of outputs, or loss of sensitive data, can be devastating. Adversaries in this domain are sophisticated, well-resourced, and strategically motivated, making the need for robust security requirements for AI systems in foreign policy all the more necessary. 

\textsc{Operationalization} translates governance methods to institutional practice \cite{reuel_open_2025}. In conducting statecraft, the operationalization gap is particularly acute due to the myriad pressures on idiosyncratic interpersonal practices. For example, hierarchy, accountability, and confidentiality concerns can distort the effective translation of evaluation outputs. As such, AI governance in foreign policy requires a deep understanding of potential deployment protocols that can preserve human authority over contestable judgments.

\textsc{Ecosystem Monitoring} refers to the ongoing study of the evolving landscape of AI development, application, and associated impacts \cite{reuel_open_2025}. AI systems used for foreign-policy use cases can inhibit such monitoring. For example, the classified nature of many government AI deployments makes it difficult to assess the degree of model monoculture, that is, the extent to which states are relying on a narrow set of systems or architectures that create systemic vulnerabilities. Compounding this, vendor dependence within government can become deeply entrenched over time, with shared data pipelines and other integrated infrastructure making it difficult to transition away from specific providers. Similarly, procurement processes in government contexts may present a high bar for new market entrants, limiting opportunities for experimentation. Foreign policy, therefore, requires longitudinal evaluation and ecosystem-level monitoring, not only static benchmark scores.

Taken together, these properties imply that foreign policy deployments cannot be governed through model-only leaderboards. Following \citeauthor{weidinger_sociotechnical_2023} \citeyear{weidinger_sociotechnical_2023}, evaluation should incorporate capability testing alongside human interaction and system-level effects –– both of which require a more rigorous investigation in statecraft deployments. As in \citeauthor{shevlane_model_2023} \citeyear{shevlane_model_2023}, high risk settings should separate the possible capabilities of a system from its disposition under realistic institutional incentives and deployment settings. These are particularly acute pressures in the conduct of statecraft and will demand more investigation on the systemic implications of combined human-AI in contested domains. 

Of the six governance subareas, \textsc{Assessment} remains the most immediate entry point for technical researchers, but foreign-policy evaluation cannot remain \textsc{Assessment}-only. The evidence map below therefore empirically evaluates whether public AI evaluation and technical governance papers focus on systems or workflows in statecraft-relevant settings. Alongside \textsc{Access}, \textsc{Assessment} forms the foundation upon which all other governance mechanisms depend. Without the ability to assess the capabilities of AI systems, achieving meaningful  oversight becomes significantly complicated, if not impossible. 

Critically, TAIG researchers are best positioned to make immediate and meaningful progress on evaluation practices that could have a widespread impact on human life. The current literature on evaluations in foreign policy and statecraft is sparse, however. Through a literature review and quantitative analysis of existing work, we find strong evidence that this domain of technical AI governance research is underdeveloped.

\section{Evidence from a Corpus Screen}

Technical AI governance has developed rapidly around the problem of making advanced AI systems measurable, before and during deployment. Despite these gains, progress has been uneven. 

\subsection{Methodology}

We report an evidence map of titles and abstracts with 93{,}684 primary rows, with 76{,}653 records with title and abstract and 3{,}089 title-only records. The study window includes works published from 2018 through 2026-04-22. 

The merged corpus draws from three overlapping source families: a main AI/ML/NLP venue corpus, an AI governance, policy, safety, assurance, and evaluation corpus, and a statecraft-adjacent bridge corpus. Because corpus memberships are not mutually exclusive, pooled results should be interpreted as merged descriptive results rather than independent corpus sums.

We first ran a high-recall screen over titles and abstracts, normalizing text and matching AI/ML terms, TAIGR method terms, and statecraft-adjacent terms. From this, we built an adjudication and negative audit pipeline, adjudicating the top 1{,}370 candidates and a 500-record negative-audit sample that would have been excluded based on the prior matching. Each row was coded for relevance to statecraft and TAIGR evaluation method relevance. 

\begin{table}[t]
\centering
\small
\begin{tabular}{lr}
\toprule
Quantity & Result \\
\midrule
Primary rows in merged warehouse & 93{,}684 \\
Headline rows screened & 79{,}954 \\
Title-and-abstract rows & 76{,}653 \\
Title-only rows & 3{,}089 \\
Non-paper accounting rows & 212 \\
LLM adjudication queue & 1{,}370 \\
Accepted domain+method rows & 413 \\
Raw direct/proxy adjudicated labels & 14 \\
Direct real-world labels before QA & 7 \\
Proxy statecraft benchmark labels & 7 \\
Negative-audit strict misses & 0 \\
\bottomrule
\end{tabular}
\caption{Current corpus-screening and adjudication evidence for the foreign-policy evaluation gap. Counts are title-and-abstract-first screening and adjudication results. The 413 accepted domain+method rows are broader contextual evidence and should not be interpreted as direct foreign-policy AI evaluation coverage.}
\label{tab:evidence-glance}
\end{table}

\subsection{Empirical Evidence}

TAIGR has built substantial machinery for a range of analyses that could be relevant to high-stakes domains like statecraft. This includes strong work on assessing and benchmarking capabilities, but also the safety, privacy, and equity implications, with policy-facing analysis on security and systemic considerations. Despite this work, the corpus screen gives little evidence of a mature evaluation agenda for foreign-policy decision contexts. The field is not empty, but the direct literature remains small and overwhelmingly \textsc{Assessment}-centered. We identify 12 papers (out of 79{,}954), approximately 0.015\% of the screened corpus. Five are direct real-world statecraft evaluations, and seven are proxy statecraft benchmarks.\footnote{Our five paper-ready direct real-world hits are: "Not Oracles of the Battlefield: Safety Considerations for AI-Based Military Decision Support Systems"; "Human vs. Machine: Behavioral Differences between Expert Humans and Language Models in Wargame Simulations"; "Escalation Risks from Language Models in Military and Diplomatic Decision-Making"; "This Land is Your, My Land: Evaluating Geopolitical Bias in Language Models through Territorial Disputes"; "DiplomacyAgent: Do LLMs Balance Interests and Ethical Principles in International Events?". The seven proxy statecraft benchmark hits include no-press and full-press Diplomacy work, Civilization/CivRealm, Cicero Diplomacy play, and DipLLM.}

The broader set of 413 domain relevant papers indicate active evaluation and governance methods in nearby areas, especially cyber/information operations and generic negotiation or mixed-motive agent evaluation. These papers, however, miss many of the central structural features of taking actions in world politics. Statecraft is not wholly absent, but directly relevant work remains scarce.

The adjudicated results and accompanying literature review reveal four methodological gaps in current evaluation approaches.

\emph{1. Evaluations often lack ecological validity.}\\
Evaluation settings for foreign policy, negotiation, and diplomacy are often deliberately simplified in ways that fail to mirror real-world complexity \cite{bakhtin_human-level_2022,Kramar2022-nx,xu2025dipllmfinetuningllmstrategic,duffy2025democratizingdiplomacyharnessevaluating,johanson2022emergentbarteringbehaviourmultiagent}. While such simplification reduces confounding factors and enables tractable measurement, it distorts the real-life nature of the settings being modeled. Utility functions are typically limited to one or two dimensions such as price or success ratios \cite{bagga_deep_2020,luo_survey_2024, fu2023improvinglanguagemodelnegotiation}, while multiple-choice benchmarks flatten the open-ended nature of real, dynamic negotiations \cite{jensen2025criticalforeignpolicydecisions}. In simulation-based experiments, critical dimensions of diplomatic success, such as the durability of terms over time, are rarely accounted for. In diplomatic contexts especially, how an agreement is reached---the sequencing of concessions, the management of impasse, the gradual building of trust---is often as consequential as the final outcome itself\cite{ma2024chatgptdonttelldo}.

\emph{2. There is a lack of publicly available, common datasets for benchmarking AI systems in the foreign-policy domain.}\\
Among the works identified in our review, public resources are concentrated in highly stylized strategic environments, such as Diplomacy or Civilization-style simulations, rather than real diplomatic settings and workflows. The Automated Negotiating Agents Competition \cite{vaccaro2026advancingainegotiationslargescale} is one of the few resources approximating a relevant subdomain, but its domain focus is on generic negotiations. This is a relevant adjacent literature, but is not counted in our corpus as directly concerning statecraft due to the absence of institutionally grounded international decision contexts.

This data gap is likely motivated by the difficulty of determining, automating, and annotating datasets for ground truth. The most relevant data for benchmarking AI systems in foreign policy is drawn from real diplomatic interactions. However, this is the data that states are least willing to disclose, given its sensitivity to national security. Without standardized evaluation resources, comparability and reproducibility between different AI systems break down, creating a barrier to understanding field-wide progress in capabilities \cite{liao2025rethinkingmodelevaluationnarrowing,biderman2026lessonstrenchesreproducibleevaluation,kapoor_ai_2025}.

\emph{3. The conduct of foreign policy is often done through highly interpersonal settings.}\\
Existing evaluation methods fail to capture that nuance. Key evaluation criteria, such as managing emotional dynamics, building trust, and demonstrating cultural sensitivity, are critical to the success of a negotiation \cite{article, Elfenbein2006-hm}, but these interpersonal dimensions are rarely evaluated \cite{ma2024chatgptdonttelldo}. By reducing diplomacy to strictly outcome metrics, existing evaluation frameworks risk encoding a transactional conception of negotiation and conflict management. As evaluations are used to inform research directions, such encoding could systematically reward the wrong behaviors for future models and mischaracterize effective foreign policy.

\emph{4. There is considerable fragmentation of evaluation approaches across studies.} \\ 
This fragmentation reflects the diversity of conceptualizations of success for different players across foreign-policy domains and contexts \cite{putnam_diplomacy_1988,allison_essence_1999}. A nation's foreign policy involves many competing goals. For instance, an agent evaluated on economic outcomes in a trade negotiation context cannot be meaningfully compared to one assessed on ceasefire durability in a conflict mediation setting. This fragmentation makes it difficult to advance a \emph{general} AI system across foreign-policy use cases, or build cumulative knowledge about holistic model performance. 

Adjacent evaluations in generic negotiations and proxy strategic games provide partial evidence, but they do not measure the same constructs. An agent evaluated on economic surplus in a generic bargaining task cannot be directly compared with a model evaluated on geopolitical bias in a statecraft analysis workflow. Evaluations should therefore distinguish domain-specific criteria, such as ceasefire analysis or sanctions compliance, from more general process-quality criteria, such as provenance, uncertainty preservation, escalation-to-review behavior, and robustness to strategic misrepresentation.

Shared methodological benchmarks should enable reproducible comparisons within well-defined task families rather than force comparability across heterogeneous statecraft-adjacent domains. Plausible scenarios grounded in real cases should be paired with evaluation criteria that account for both process and outcome quality \cite{schwartz2025realitychecknewevaluation}. As AI systems become more widely used in foreign-policy contexts, evaluation practices will need to match the complexity of the environments they are meant to govern.

\section{Towards A Demand-Side Evaluation Agenda}

Our corpus scan suggests a supply-side bias. Much of the broader accepted literature begins with a model and failure mode, then selecting tasks to expose it. The statecraft-relevant strict hits are too sparse and heterogeneous to support a model-only leaderboard. AI governance in foreign-policy settings should instead start from the demand side. This involves understanding the workflows that need supporting and testing AI systems within the institutional and organizational structures that give authority and accountability.

\Cref{tab:demand-side} gives an illustrative set of task families to shift evaluation design away from generic model leaderboards and toward task-scoped tests that map onto real foreign-policy workflows.

\begin{table*}[t]
\centering
\small
\begin{tabular}{p{.11\textwidth}p{.17\textwidth}p{.20\textwidth}p{.25\textwidth}p{.18\textwidth}}
\toprule
Stage & Task & Output & Evaluation & Primary TAIG capacities \\
\midrule
Research & Actor and constraint mapping & Actor-interest-constraint map & Coverage, source provenance, uncertainty marking, omission detection & Assessment, Verification \\
Analyze & Escalation-signal detection & Risk assessment with evidence links & Calibration, adversarial robustness, false-positive control, caveat preservation & Assessment, Security \\
Strategize & Option generation & Menu of policy options & Trade-off coverage, legal feasibility, non-escalatory alternatives, red-team objections & Assessment, Operationalization \\
Execute & Draft-language review & Flagged clauses, caveats, obligations & Citation fidelity, ambiguity handling, obligation tracking, human-review triggers & Verification, Operationalization \\
Monitor & Agreement compliance tracking & Compliance assessment and unresolved issues & Provenance, longitudinal consistency, source conflict handling, escalation to review & Verification, Ecosystem Monitoring \\
\bottomrule
\end{tabular}
\caption{Demand-side evaluation agenda for foreign-policy AI systems. The unit of evaluation is task fitness under institutional and strategic constraints, evaluated through bounded artifacts rather than generic model performance.}
\label{tab:demand-side}
\end{table*}

For each task family, researchers should build task cards: structured specifications that define the task-relevant information. A task card for escalation detection, for example, should specify the scenario, signals, distortions, and the conditions of system uncertainty.

For a demand-side evaluation, researchers should specify discrete tasks, the human role, and potential scoring regimes. Accurately characterizing the task means bounding the activity appropriately within a complex existing system, with the model's role and definition of success. This might include mapping actors and their positions, or generating policy options. The human role should also be made clear –– at each point in these workflows, responsibility should return to an accountable human decision-maker. Any scoring regime would need to specify what can be evaluated directly, what must be judged by experts, and what should be treated as unresolvable within the benchmark.
         
A useful evaluation may ask whether a system faithfully accurately cited evidence or designed strategies that obeyed certain laws. These properties are more tractable than evaluating whether the model ``solved'' a foreign-policy problem or exhibited some disposition or proclivity during strategic reasoning.

Good demand-side evaluations should also be able to compare systems against real baselines. Models may perform well in isolation, but comparison between human--AI workflows and the pre-existing institutional processes will be important evaluation. Understanding relative strengths and weaknesses, as well as failure modes, is easier with strong baselines. Where possible, comparisons between human-only performance, model-only performance, and human-model collaboration should help trace the evolution of (over)reliance and the state of capacities for judgment, both human and institutional.

Finally, foreign-policy evaluation resources require strict governance. Public release should be encouraged, but sensitive evaluations that explore vulnerabilities might want to consider controlled-access settings, with expert review panels or secure audit sandboxes. Though the cumulative improvement of evaluation science is important, the unrestricted publication of vulnerabilities could weaken the relative position of more responsible parties. Demand-side evaluation is therefore both a technical agenda and an access-design problem: the field needs benchmarks, but it also needs governance arrangements that determine who can run them and when information needs to be disclosed. 

\subsection{Limitations}

Our claims are limited to the public, externally inspectable research ecosystem, and comes with several potential limitations. First, the public literature gap may overstate the practical governance gap because classified evaluation practices may exist inside governments or vendors. Second, our main screening run is based on titles and abstracts alone and doesn't consider full-text content. Third, critics may argue that foreign policy should not be benchmarked at all, noting that benchmark construction bears the risk of flattening contested political judgment and detaching human agency from key normative issues. Finally, foreign-policy benchmarks may be dual-use: better evaluation can improve safety, but it can also improve systems used for coercion, deception, or strategic manipulation.

We agree that model-only leaderboards for holistic diplomatic activities would not be a useful contribution. Our proposal is scoped, task-based evaluation infrastructure for human-supervised institutional workflows.

\section{Conclusion}

Foreign policy and statecraft settings should be a priority domain for technical AI governance because it combines stark consequences with weak evaluability. The structures of foreign policy decisions, with private information between adversaries and contested objectives on all sides, make standard benchmark practice insufficient. The resulting governance gap needs addressing: AI systems are being explored for foreign-policy workflows before the field has built evaluation infrastructure suited to those workflows.

This paper argues for a demand-side research agenda. Instead of treating foreign policy as a single benchmarkable capability, evaluations should break down institutional workflows into decomposed, bounded tasks. These tasks should define the downstream outputs, preserving human authority over judgment and accountability over outcomes. The agenda outlined in this paper gives technical AI governance a tractable path into a domain where disciplined evaluation is urgently needed.

\newpage

\section*{Impact Statement}

This paper does not develop, deploy, or advocate any AI system for foreign policy. Its primary impact is to shape how such systems are evaluated if governments, international organizations, or vendors adopt them. Better evaluation could reduce unsafe deployment, reveal failure modes, and improve institutional accountability. It could also improve systems used for coercive or destabilizing state action. For this reason, we emphasize task-scoped evaluation, practitioner validation, controlled access, auditability, and accountable human authority rather than unrestricted benchmark release or automated diplomacy.

Evaluation infrastructure is not neutral. Benchmarks and scoring regimes can encode narrow conceptions of success, especially in domains shaped by unequal power, secrecy, and contested legitimacy. The paper therefore treats evaluation as a governance problem as well as a technical problem: evaluations should specify whose objectives are represented, what forms of harm are considered, what information is excluded, and where human responsibility remains non-delegable.

\section*{LLM Usage Statement}

During the preparation of this work, the authors used Claude by Anthropic to improve readability and formatting of the manuscript. ChatGPT, Claude, and OpenAI Codex were also used to support corpus-analysis workflows, including code assistance, report generation, and title-abstract adjudication. The adjudication procedure used a fixed schema and conservative boundary rules. This involved hand-calibrated labels for all strict core rows, as well as the first 50 papers from cyber/info-ops, QA-checked, duplicate work IDs, strict hits, and the negative-audit candidates. The authors reviewed the outputs used in the manuscript and take full responsibility for the final content.

\bibliography{refs.bib}
\bibliographystyle{icml2026}

\newpage
\appendix
\onecolumn
\section{Appendix: Corpus Screening and Codex-Adjudicated Evidence Map}
\label{app:corpus-screen}

This appendix documents the corpus-screening and adjudication procedure supporting Section~5. This screen estimates whether public technical AI governance research substantially covers AI systems or AI-enabled workflows in foreign-policy and statecraft contexts through a title-and-abstract evidence map over a bounded public research corpus.

\subsection{Purpose and Scope}
\label{app:purpose-scope}

The study looks at whether public TAIGR covers statecraft-relevant settings (e.g., foreign policy, diplomacy, military decision-making, intelligence workflows, crisis management, arms control, sanctions, information operations, cyber operations, geopolitical governance, and state-to-state strategic interaction).

A substantive hit requires \emph{both} foreign policy domain relevance \emph{and} TAIGR relevance. The title and abstract must indicate that the paper evaluates, benchmarks, audits, governs, verifies, monitors, red-teams, assesses, or operationalizes an AI system or AI-enabled workflow in a relevant domain. Full text was not used as part of the primary screen.

\subsection{Corpus}
\label{app:corpus-denominator}

The study window includes works with a published date from 2018-01-01 through 2026-04-22. Where only publication year was available, works from 2018 through 2026 were included and treated as year-only window rows. Pre-window, post-window, and unknown-year rows were excluded from the headline denominator.

The merged warehouse contains 93,684 primary rows. The study-window slice used by the deterministic screen contained 79,954 rows. This denominator includes 76,653 title-and-abstract rows, 3,089 title-only rows, and 212 non-paper accounting rows retained. 

The merged corpus consolidates three overlapping source families: (i) a main AI/ML/NLP venue corpus, (ii) an AI governance, policy, safety, assurance, and evaluation corpus, and (iii) a statecraft-adjacent bridge corpus. Corpus memberships are not mutually exclusive. Pooled results should therefore be interpreted as merged descriptive results, not independent corpus sums.

\begin{table}[h]
\centering
\small
\begin{tabular}{lr}
\toprule
Quantity & Result \\
\midrule
Primary rows in merged warehouse & 93,684 \\
Headline rows screened & 79,954 \\
Title-and-abstract rows & 76,653 \\
Title-only rows & 3,089 \\
Non-paper accounting rows & 212 \\
LLM/Codex adjudication queue & 1,370 \\
Accepted domain+method rows & 413 \\
Raw direct/proxy adjudicated labels & 14 \\
Conservative paper-ready strict hits & 12 \\
Direct real-world paper-ready hits & 5 \\
Proxy statecraft benchmark paper-ready hits & 7 \\
Negative-audit strict misses & 0 \\
\bottomrule
\end{tabular}
\caption{Current corpus-screening and adjudication evidence. Counts are title-and-abstract-first screening and adjudication results. The 413 accepted domain+method rows are broader contextual evidence and should not be interpreted as direct foreign-policy AI evaluation coverage.}
\label{tab:appendix-current-evidence}
\end{table}

\subsection{Deterministic Screen}
\label{app:deterministic-screen}

The first stage was a deterministic screen over titles and abstracts. For each work, evidence text was built from the title and abstract, normalized for punctuation, and matched to four families of terms: AI/ML, TAIGR evaluation/governance methods, statecraft-relevant, and statecraft-adjacent (e.g., cyber, information-operations, geopolitical-governance, and generic-negotiation). This deterministic screen built the queue for deeper screening, rows outside likely-hit buckets formed the negative-audit pool.

\begin{table}[h]
\centering
\small
\begin{tabular}{lr}
\toprule
Deterministic bucket & Rows \\
\midrule
Negative audit pool & 52,477 \\
Exclude: method only & 24,629 \\
Exclude: domain only & 1,677 \\
Generic negotiation or agent evaluation & 768 \\
Adjacent cyber/info-ops evaluation & 354 \\
Adjacent geopolitical AI governance & 23 \\
Sparse statecraft-AI review & 10 \\
Core direct statecraft evaluation & 9 \\
Core proxy statecraft evaluation & 7 \\
\bottomrule
\end{tabular}
\caption{Deterministic-screen bucket counts over the 79,954-row headline denominator. The screen was designed for recall and did not determine final inclusion.}
\label{tab:deterministic-buckets}
\end{table}

A known-positive recall check used 17 seed papers. All 17 were found by the deterministic screen; none were missed by regex and none were absent from the denominator.

\subsection{LLM Adjudication Queue}
\label{app:codex-queue}

The adjudication queue contained 1,370 rows. It included all strict/core candidates, adjacent cyber and information-operations candidates, adjacent geopolitical and sparse-review candidates, generic negotiation and agent candidates, and a 500-row negative-audit sample drawn from deterministic negatives.

\begin{table}[h]
\centering
\small
\begin{tabular}{lr}
\toprule
Queue source & Rows \\
\midrule
Negative audit pool sample & 500 \\
Generic negotiation or agent evaluation & 472 \\
Adjacent cyber/info-ops evaluation & 354 \\
Adjacent geopolitical AI governance & 20 \\
Core direct statecraft evaluation & 9 \\
Sparse statecraft-AI review & 8 \\
Core proxy statecraft evaluation & 7 \\
\midrule
Total & 1,370 \\
\bottomrule
\end{tabular}
\caption{Composition of the Codex adjudication queue. The queue deliberately does not include all 79,954 headline rows. It includes likely hits plus a negative-audit sample.}
\label{tab:codex-queue}
\end{table}

Each queued row was presented with its work identifier, title, abstract where available, venue, venue family, year, corpus membership, evidence basis, deterministic bucket, matched regex terms, and screen reason.

\subsection{Coding Schema}
\label{app:coding-schema}

Each adjudicated row received labels for statecraft relevance, TAIGR evaluation relevance, deployment context, AI object, evaluation method, confidence, an evidence quote, and notes.

The boundary was relatively conservative, papers describing deployment alone were sufficient. The title or abstract had to indicate some form of TAIGR capacity.

\subsection{Adjudicated Results}
\label{app:adjudicated-results}

The LLM (GPT-5.5) adjudication accepted 413 rows as domain+method relevant. Most accepted rows were from adjacent domains rather than direct statecraft. After adjudication, the conservative strict count was 12: five direct real-world rows and seven proxy statecraft-benchmark rows. Two additional labeled papers were deemed borderline due to limited evaluation investigation.

\begin{table}[h]
\centering
\small
\begin{tabular}{lrrr}
\toprule
Label & Rows & Share of queue & Share of denominator \\
\midrule
Direct real-world & 7 & 0.51\% & 0.009\% \\
Proxy statecraft benchmark & 7 & 0.51\% & 0.009\% \\
Adjacent geopolitical governance & 4 & 0.29\% & 0.005\% \\
Adjacent cyber/info-ops & 310 & 22.63\% & 0.388\% \\
Generic negotiation/mixed-motive & 85 & 6.20\% & 0.106\% \\
Rejected / not relevant & 957 & 69.85\% & 1.197\% \\
\midrule
Total & 1,370 & 100.00\% & 1.713\% \\
\bottomrule
\end{tabular}
\caption{Codex-adjudicated relevance labels (pre-QA counts). Manual QA subsequently reduces the direct real-world count from 7 to 5 (see the strict direct/proxy QA table); the conservative post-QA strict total is 12. Shares of the denominator use the 79,954-row headline denominator.}
\label{tab:codex-label-counts}
\end{table}

\begin{figure}[h]
\centering
\includegraphics[width=.85\linewidth]{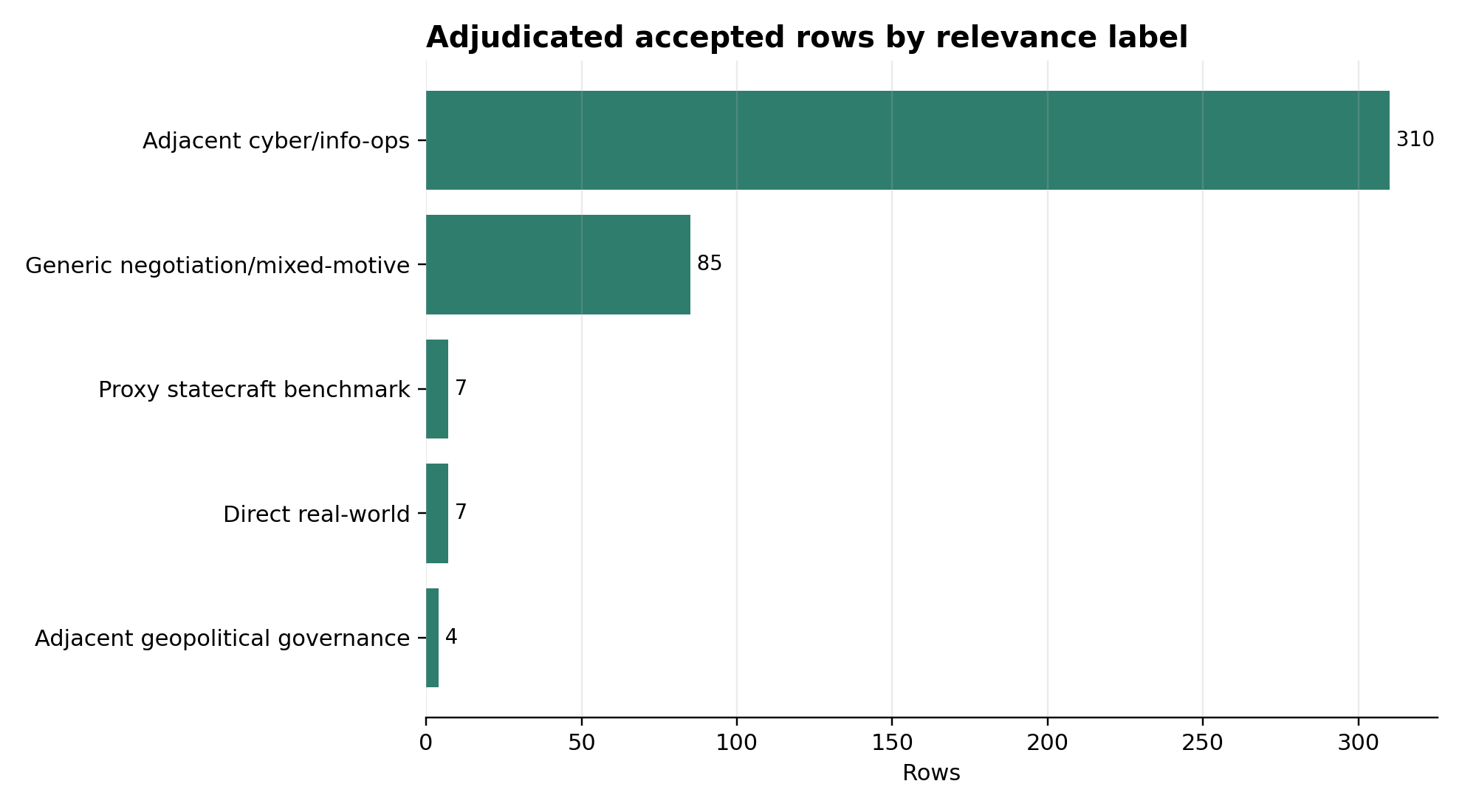}
\caption{Adjudicated accepted rows by relevance label.}
\label{fig:adjudicated-relevance-counts}
\end{figure}

\begin{figure}[h]
\centering
\includegraphics[width=.85\linewidth]{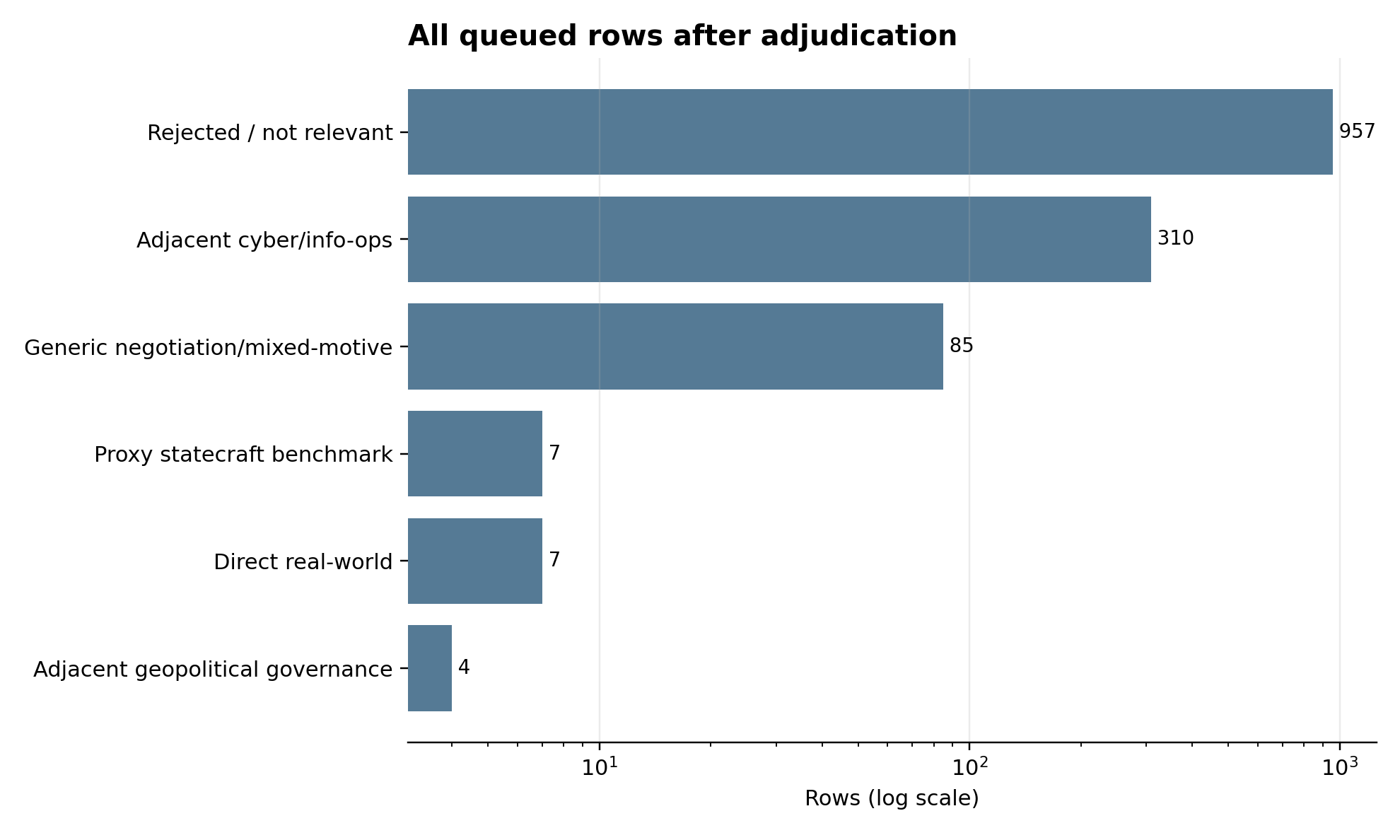}
\caption{All adjudicated labels, including rejected or not-relevant rows.}
\label{fig:all-adjudicated-labels}
\end{figure}

\subsection{Strict Direct and Proxy QA}
\label{app:strict-hit-qa}

The raw direct/proxy adjudicated count was 14. Manual QA retained 12 rows as conservative paper-ready strict evidence and marked two direct-labeled rows as contextual or borderline. The conservative strict count is the headline count used in the paper.

\begin{table*}[t]
\centering
\small
\begin{tabular}{p{0.47\linewidth}rlp{0.19\linewidth}}
\toprule
Title & Year & Venue & QA status \\
\midrule
Not Oracles of the Battlefield: Safety Considerations for AI-Based Military Decision Support Systems & 2024 & AIES & Paper-ready strict \\
Human vs. Machine: Behavioral Differences between Expert Humans and Language Models in Wargame Simulations & 2024 & AIES & Paper-ready strict \\
Escalation Risks from Language Models in Military and Diplomatic Decision-Making & 2024 & FAccT & Paper-ready strict \\
This Land is Your, My Land: Evaluating Geopolitical Bias in Language Models through Territorial Disputes & 2024 & NAACL & Paper-ready strict \\
DiplomacyAgent: Do LLMs Balance Interests and Ethical Principles in International Events? & 2025 & EMNLP & Paper-ready strict \\
Human-Level Performance in No-Press Diplomacy via Equilibrium Search & 2021 & ICLR & Paper-ready strict \\
No-Press Diplomacy from Scratch & 2021 & NeurIPS & Paper-ready strict \\
Mastering the Game of No-Press Diplomacy via Human-Regularized Reinforcement Learning and Planning & 2023 & ICLR & Paper-ready strict \\
CivRealm: A Learning and Reasoning Odyssey in Civilization for Decision-Making Agents & 2024 & ICLR & Paper-ready strict \\
More Victories, Less Cooperation: Assessing Cicero's Diplomacy Play & 2024 & ACL & Paper-ready strict \\
DipLLM: Fine-Tuning LLM for Strategic Decision-making in Diplomacy & 2025 & ICML & Paper-ready strict \\
Democratizing Diplomacy: A Harness for Evaluating Any Large Language Model on Full-Press Diplomacy & 2026 & AAAI & Paper-ready strict \\
\midrule
Knowledge-Embedded Narrative Construction from Open Source Intelligence & 2023 & AAAI & Borderline/contextual \\
A Survey of Large Language Model Use and Its Technical Limitations in Military Systems Through a Decolonial Lens & 2025 & AIES & Borderline/contextual \\
\bottomrule
\end{tabular}
\caption{Strict direct/proxy QA. Twelve rows are retained as conservative paper-ready strict evidence. Two additional direct-labeled rows are retained as contextual or borderline and excluded from the headline strict count.}
\label{tab:strict-hit-qa}
\end{table*}

The two contextual rows were excluded from the conservative headline for two reasons. \emph{Knowledge-Embedded Narrative Construction from Open Source Intelligence} is intelligence-workflow relevant, but is more of a proposed system rather than an evaluation methodology. \emph{A Survey of Large Language Model Use and Its Technical Limitations in Military Systems Through a Decolonial Lens} is relevant to military LLM use, but does not provide a concrete evaluation, benchmark, audit, or workflow test of a deployed or simulated statecraft process.

The negative audit sampled 500 rows from the deterministic negative pool. The audit produced four possible false negatives. All four were low-confidence adjacent cyber/info-ops rows. None were paper-ready strict direct or proxy misses.

\begin{table*}[t]
\centering
\small
\begin{tabular}{rp{0.62\linewidth}ll}
\toprule
Work ID & Title & Label & Confidence \\
\midrule
56454 & DevFD: Developmental Face Forgery Detection by Learning Shared and Orthogonal LoRA Subspaces & Adjacent cyber/info-ops & Low \\
10067 & IOHunter: Graph Foundation Model to Uncover Online Information Operations & Adjacent cyber/info-ops & Low \\
32651 & PPT-GNN: A Practical Pretrained Spatio-Temporal Graph Neural Network for Network Security & Adjacent cyber/info-ops & Low \\
5412 & Tracking and Identifying International Propaganda and Influence Networks Online & Adjacent cyber/info-ops & Low \\
\bottomrule
\end{tabular}
\caption{Possible false negatives from the 500-row negative audit. All four are low-confidence adjacent cyber/info-ops rows; none are strict direct or proxy statecraft misses.}
\label{tab:negative-audit}
\end{table*}

The negative audit supports the stability of the conservative statecraft claim. Improving recall for adjacent cyber/info-ops would require broader regex terms, but the audit did not reveal missed paper-ready direct real-world or proxy statecraft-benchmark evidence.

\begin{figure}[h]
\centering
\includegraphics[width=.85\linewidth]{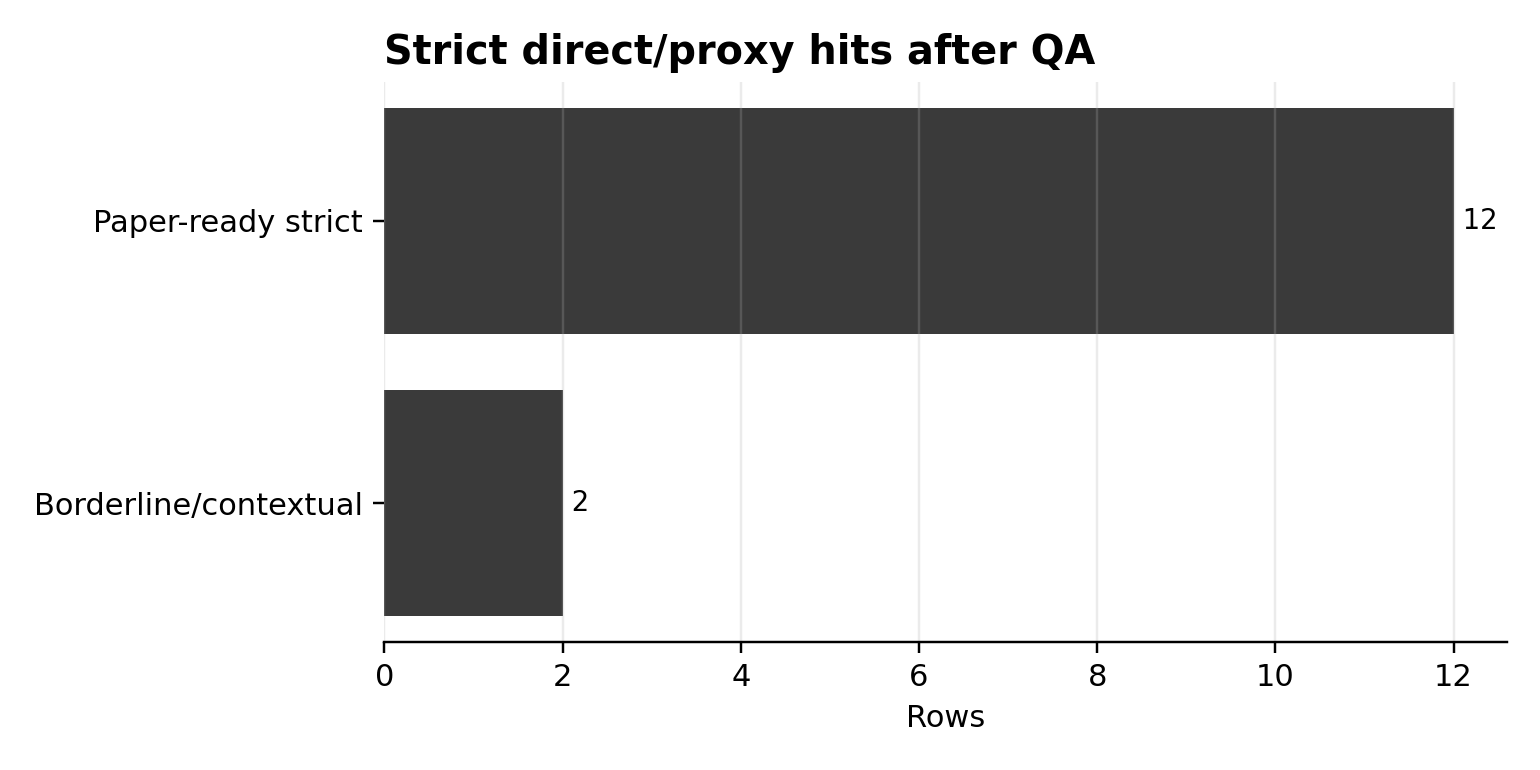}
\caption{Manual QA split between paper-ready strict and borderline/contextual direct or proxy rows.}
\label{fig:strict-hit-qa}
\end{figure}

\subsection{Methods Among Accepted Rows}
\label{app:accepted-methods}

The 413 domain+method papers included evaluations, benchmarks, datasets, tools, and platforms. This broader set is useful for locating nearby methodological activity, but it should not be collapsed into direct statecraft evidence.

\begin{table}[h]
\centering
\small
\begin{tabular}{lr}
\toprule
Method & Rows \\
\midrule
Benchmark/dataset & 202 \\
Tool/platform & 42 \\
Empirical model evaluation & 40 \\
Red team/adversarial & 39 \\
Survey/taxonomy/position & 32 \\
Human/expert study & 30 \\
Policy/legal analysis & 27 \\
Other/unclear & 1 \\
\midrule
Total & 413 \\
\bottomrule
\end{tabular}
\caption{Evaluation and governance methods among the 413 accepted domain+method rows.}
\label{tab:accepted-methods}
\end{table}

\begin{figure}[h]
\centering
\includegraphics[width=.85\linewidth]{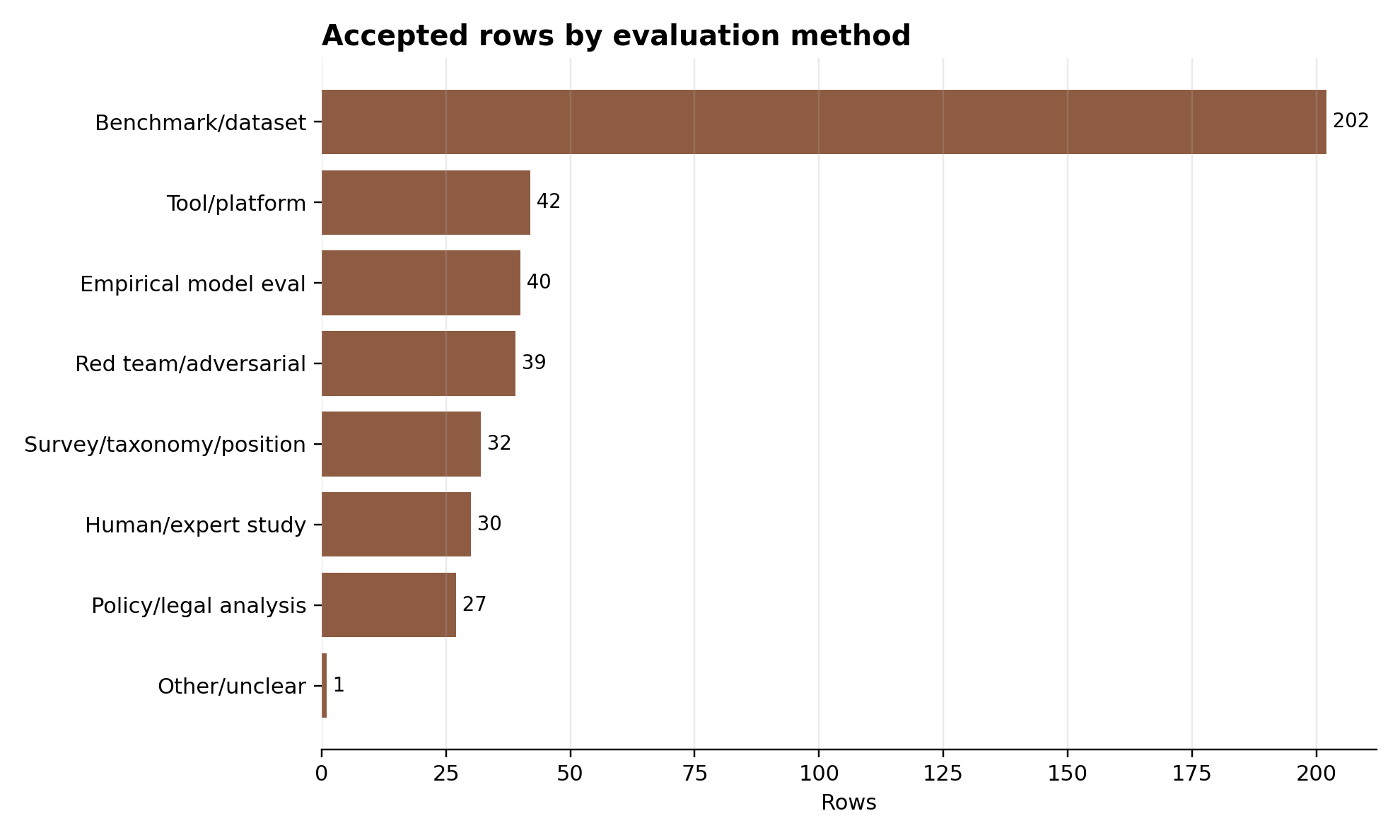}
\caption{Accepted rows by evaluation or governance method.}
\label{fig:accepted-by-method}
\end{figure}

\begin{figure}[h]
\centering
\includegraphics[width=.95\linewidth]{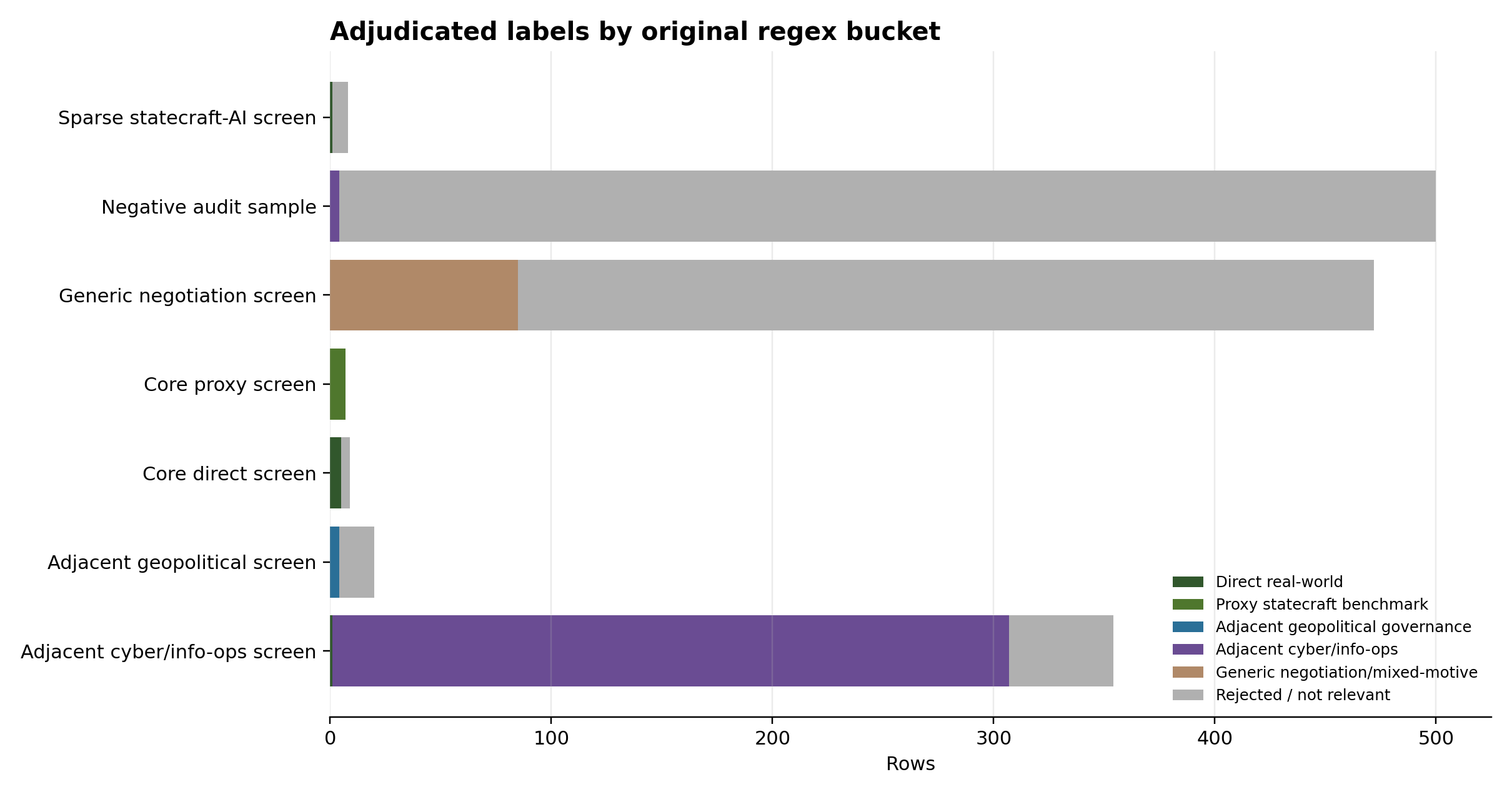}
\caption{Movement from deterministic candidate buckets to final adjudicated labels.}
\label{fig:bucket-to-label-stacked}
\end{figure}

\subsection{Reproducibility}
\label{app:reproducibility}
The row-level files accompanying this appendix preserve paper identifiers, evidence-basis fields, confidence labels, and high-level categories for domain and TAIG capacity. The aggregate CSVs include the tables reported above and additional domain-by-year, venue-by-year, domain-by-capacity, and domain-by-method cross-tabulations. The figure files used in this ICML-formatted appendix are included in the figures/ directory.


\end{document}